\documentclass[a4paper,12pt]{article}
\usepackage{rotating,threeparttable,booktabs}
\usepackage{epsfig}
\usepackage{psfrag,graphicx}
\usepackage{subfigure}
\begin{document}
\renewcommand{\baselinestretch}{01.35}
\renewcommand{\arraystretch}{0.666666666}
%\parindent=0pt
%{\large
\parskip.2in
\newcommand{\hs}{\hspace{1mm}}
\newcommand{\nhat}{\mbox{\boldmath$\hat n$}}
\newcommand{\cmod}[1]{ \vert #1 \vert ^2 }
\newcommand{\mod}[1]{ \vert #1 \vert }
\newcommand{\pr}{\partial}
\newcommand{\fr}{\frac}
\newcommand{\ie}{{\em ie }}
\newcommand{\p}{\varphi}
\newcommand{\xib}{{\bar{\xi}}}
\newcommand{\Vd}{V^{\dagger}}
\newcommand{\Ref}[1]{(\ref{#1})}
\title{{\Large{\bf Skyrmion Vibration Modes within the Rational Map Ansatz}}}

\author{
W.T. Lin\thanks{e-mail address: wen-tsan.lin@durham.ac.uk}\, and
B. Piette\thanks{e-mail address: B.M.A.G.Piette@durham.ac.uk}
\\ Department of Mathematical Sciences,University of Durham, \\
 Durham DH1 3LE, UK\\
}
\date{}

\maketitle

\begin{abstract}
We study the vibration modes of the Skyrme model within the rational map 
ansatz. We show that the vibrations of the radial profiles and the rational 
maps are decoupled and we consider explicitly the case $B=1$, $B=2$ and $B=4$.
We then compare our results with the vibration modes obtained numerically by 
Barnes et al. and show that qualitatively the rational map reproduces 
the vibration modes obtained numerically but that the vibration frequencies
of these modes do not match very well. 
\end{abstract}

%%%%%%%%%%%%%%%%%%%%%%%%%%%%%%%%%%%%%%%%%%%%%%%%%%%%%%%%%%%%%%%%%%%%%%%%%%%%
\section{Introduction}
Proposed by Skyrme as a fundamental theory of strong interactions,
the Skyrme\cite{Skyrme} model was later shown by Witten\cite{Witten} 
to be a low energy limit of QCD in the limit of large colour number.
In that context, the classical solutions of the Skyrme model correspond to 
bound states of QCD. The simplest solution, with baryon number 1, 
can be computed analytically up to solving an ordinary 
differential equation 
numerically. For larger baryon numbers, one must compute the solutions 
by solving the full classical equation of the model numerically\cite{BS1}. 

Recently, Houghton et al.\cite{HMS} showed that the solution of the Skyrme 
model can be well approximated by the so called Rational Map ansatz.
In this ansatz solutions are approximated by a radial profile function and a
rational map ansatz which only depends on the polar angles variables. One then
determines the radial profile by solving an ordinary differential equation 
while the rational map minimises an integral defined on the sphere.
The configurations obtained by the ansatz fit the numerical solutions very 
well.
  
Once a classical solution has been obtained, one must still quantise some of 
the remaining degree of freedom. One way to do this is to compute the vibration 
modes of these solutions. This was done for the numerical solutions of the 
model by Barnes et al.\cite{BBT1}\cite{BBT2} for baryon numbers $B=2$ and $4$. 
In this paper
we compute the vibrational mode of the rational map configuration and
compare them to those obtained numerically. 

The Skyrme Model\cite{Skyrme} is defined by the following Lagrangian
\begin{eqnarray}
L_p&=&\int_{R^3}\bigg\{\,\frac{F_{\pi}^2}{16}\,
Tr(\partial_{\mu}U\,\partial^{\mu}U^{\dag})+\frac{1}{32e^2}\,
Tr([(\partial_{\mu}U)\,U^{\dag}\,,\,(\partial_{\nu}U)U^{\dag}]^2)\nonumber\\
&&
+\frac{F_{\pi}^2\,m_{\pi}^2}{8p^2}\,Tr(U^p-\mathbf{1})\bigg\}\,d^3x\;,
\label{dimlag}
\end{eqnarray}
where $U=U(\vec{x},t)$ is an $SU(2)$ chiral field, $F_\pi$ is the pion 
decay constant, $m_\pi$ is the pion mass and $e$ a parameter of the model
which is determined by fitting the classical solutions to experimental data.
Notice that the last term in (\ref{dimlag}) is the so called mass term where
we have used the generalised mass term proposed by Kopeliovich et 
al.\cite{KPZ} where the parameter $p$ is a positive integer.

Rather than using (\ref{dimlag}), it is convenient to rescale the space-time 
coordinates and the mass parameter, 
${\tilde x_\mu} = 2 x_\mu /F_\pi e; m = m_\pi 2/F_\pi e$ and use the 
dimensionless Lagrangian
\begin{eqnarray}
L&=&\frac{e}{3\pi^2F_{\pi}}\,L_p\nonumber\\
&=&\frac{1}{12\pi^2}\,\int\bigg\{-\frac{1}{2}Tr(R_{\mu}R^{\mu})
+\frac{1}{16}Tr\Big(\big[R_{\mu},R_{\nu}\big]\big[R^{\mu},R^{\nu}\big]\Big)
+\frac{m^2}{p^2}Tr\big(U^p-\mathbf{1}\big)\bigg\}\,d^3x\;,
\label{lag}
\end{eqnarray}
where $R_{\mu} = (\partial_{\mu}U) U^\dag$.

To approximate the solution by the rational map ansatz, we first introduce
the complex coordinate $\xi = \tan(\frac{\theta}{2})\,e^{i\phi}$ where $\theta$
and $\phi$ are the polar angles. Then the rational map ansatz is given by
\begin{eqnarray}\label{E2.15}
U=e^{2if(r)\,\hat{n}_{R(\xi)}\,\cdot\,\vec{\sigma}}\;,
\label{RMansatz}
\end{eqnarray}
where $\vec{\sigma}$ are the Pauli matrices and
\begin{equation}
\hat{n}_{R(\xi)}=\frac{1}{1+|R(\xi)|^{\,2}}\,\Big(2\Re(R(\xi)),
2\Im(R(\xi)), 1-|R(\xi)|^{\,2}\Big)\;.
\end{equation}
The degree of the rational map $R(\xi)$ corresponds to the baryon number of 
the configuration (see \cite{HMS}).

To approximate the classical solution of a given baryon number B, 
one would thus take
\begin{equation}
R(\xi) = \frac{P(\xi)}{Q(\xi)} 
       = \frac{\sum_{i=0}^{i=B} a_i \xi^i}{\sum_{j=0}^{j=B} b_j \xi^j}
\end{equation}
and insert the ansatz (\ref{RMansatz}) into the Lagrangian (\ref{lag}).
One must first determine the parameters $a_i$ and $b_j$ which minimises the 
integral 
\begin{eqnarray}
\mathcal{I}&=&\frac{1}{4\pi}\int\Big(\frac{1+|\xi|^2}{1+|R|^2}\Big)^4\;
 \Big|\frac{dR}{d\xi}\Big|^{\,4}\;\frac{2i\,d\xi\,\bar{d\xi}}{(1+|\xi|^2)^2}\;.
\label{I}
\end{eqnarray}
Knowing the value of $\mathcal{I}$, one then uses the
Euler-Lagrange equation to derive the equation that the profile $f(r)$ must 
solve. In \cite{HMS}, it was shown that the case $B=1$ is 
nothing but the hedgehog solution computed by Skyrme. For $B=2$, the 
configuration
is axially symmetric while for $B=4$ it has the symmetry of a cube. 

\section{Vibration Modes}
To study the vibration modes of the rational map ansatz configurations 
minimising (\ref{lag}), we add a time dependant perturbation to the rational 
map and the profile function around their minimising values. 
We then insert the perturbed ansatz into the Lagrangian (\ref{lag}) and 
compute the Euler-Lagrange equation for the perturbation, keeping only the 
linear terms.

Denoting $f_0(r)$ the
minimising profile function for the static solution, we take
\begin{eqnarray}
f(r,t) = f_0(r) + g(r,t)
\label{pertF}
\end{eqnarray}
where $g$ is assumed to be a small fluctuation around $f_0$ satisfying the 
boundary condition $g(0,t) = g(\infty,t) = 0$.

To perturb a rational map, we must perturb all its coefficients,
even the one that are null, by adding a small time dependant perturbation.
\begin{eqnarray}
R(\xi,t) = P(\xi,t)/Q(\xi,t) 
\label{pertRM}
\end{eqnarray}
where
\begin{eqnarray}
P(\xi,t) &= \sum_{i=0}^{i=B} (a_i+\delta a_i) \xi^i = P_0(\xi) + \delta P(\xi,t)
\nonumber\\
Q(\xi,t) &= \sum_{j=0}^{j=B} (b_j+\delta b_j) \xi^i = Q_0(\xi) + \delta Q(\xi,t)
\end{eqnarray}
Notice that, as $R$ is a ratio of $P$ and $Q$, the coefficients of the 
rational map are determined up to an overall constant. If $a_B$ is non 
zero, we can divide both $P$ and $Q$ by $a_B(1+\delta a_B/a_B)$ to linear order
and $\delta a_B$ then can be incorporated into the other $\delta a_i$ and 
$\delta b_i$; (if $a_B$ is null, one divides $P$ and $Q$ by 
$b_B(1+\delta b_B/b_B)$). The rational map perturbations are thus 
described by $4(B+1)-2$ parameters.

When inserting (\ref{pertRM}) into (\ref{lag}), the perturbed rational map
only occurs in the integral (\ref{I}) and in the two expressions
\begin{eqnarray}
\mathcal{X}&=&\frac{1}{4\pi}\int\frac{|\dot{R}|^2}{(1+|R|^2)^2}
    \frac{2id\xi d\bar{\xi}}{(1+|\xi|^2)^2}\;,\nonumber\\
\mathcal{Y}&=&\frac{1}{4\pi}\int\frac{(1+|\xi|^2)^2}{(1+|R|^2)^4}
    \Big|\frac{dR}{d\xi}\Big|^2|\dot{R}|^2
    \frac{2id\xi d\bar{\xi}}{(1+|\xi|^2)^2}\;.
\label{XYint}
\end{eqnarray}
Defining
\begin{eqnarray}
\alpha_0&=& |P_0|^2+|Q_0|^2\nonumber\\
\alpha_1&=& P_0(\delta\bar{P})+\bar{P}_0(\delta P)+Q_0(\delta\bar{Q})
         +\bar{Q}_0(\delta Q)\nonumber\\
\alpha_2&=& (\delta P)(\delta\bar{P})+(\delta Q)(\delta\bar{Q})\\
\beta_0&=& P_{0,\,\xi}Q_0-P_0Q_{0,\,\xi}\nonumber\\
\beta_1&=& P_{0,\,\xi}(\delta Q)-P_0(\delta Q)_{\xi}-Q_{0,\,\xi}(\delta P)
        +Q_0(\delta P)_{\xi}\nonumber\\
\beta_2&=& (\delta P)_{\xi}(\delta Q)-(\delta P)(\delta Q)_{\xi}\nonumber\\
\gamma_0&=& |\beta_0|^4\nonumber\\
\gamma_1&=& 2\,|\beta_0|^2(\beta_0\bar{\beta_1}+\bar{\beta_0}\beta_1)\nonumber\\
\gamma_2&=& 4\,|\beta_0|^2\,|\beta_1|^2+(\beta_0^2\bar{\beta}_1^2+
     \bar{\beta}_0^2\beta_1^2)+2\,|\beta_0|^2(\beta_0\bar{\beta_2}
     +\bar{\beta_0}\beta_2),\nonumber\\
\lambda_1&=& Q_0(\delta\dot{P})-P_0(\delta\dot{Q})
\label{deltaPQ}
\end{eqnarray}
we have
\begin{eqnarray}
\mathcal{I}&=&\frac{1}{4\pi}\int\frac{(1+|\xi|^2)^2}{\alpha_0^4}\,
 \bigg[\gamma_0+\Big(\gamma_1-4\gamma_0\frac{\alpha_1}{\alpha_0}\Big)
\nonumber\\
 &&\;\;+\Big(10\gamma_0\frac{\alpha_1^2}{\alpha_0^2}-4\gamma_0
  \frac{\alpha_2}{\alpha_0}-4\gamma_1\frac{\alpha_1}{\alpha_0}
   +\gamma_2\Big) \bigg]\;2i\,d\xi\,\bar{d\xi}\;,\nonumber\\
\mathcal{X}&=&\frac{1}{4\pi}\int\frac{|\lambda_1|^2}{\alpha_0^2}\frac{2i\,
 d\xi\,\bar{d\xi}}{(1+|\xi|^2)^2}\;,\nonumber\\
\mathcal{Y}&=&\frac{1}{4\pi}\int
   \frac{|\beta_0|^2|\lambda_1|^2}{\alpha_0^4}\;2i\,d\xi\,\bar{d\xi}\;,
\label{IXY}
\end{eqnarray}
Notice that the integrals (\ref{I}) and (\ref{IXY}) are at most quadratic 
in the parameters 
$\delta a$ and $\delta b$. To rewrite these integrals in matrix form,
we define 
\begin{eqnarray}
\overrightarrow V(t)\equiv\left\{\begin{array}{c} \delta a_i \\
\delta b_i
\end{array} \right\}
=\left\{\begin{array}{c} \delta a_0 \\ \vdots \\ \delta a_{\mathcal{B}} \\
\delta b_0 \\ \vdots \\ \delta b_{\mathcal{B}}
\end{array} \right\}\;\;\;\;\;;\;\;\;\;\;
\dot{\overrightarrow{V}}(t)\equiv\left\{\begin{array}{c} \dot{\delta a_i} \\
\dot{\delta b_i}
\end{array} \right\}
=\left\{\begin{array}{c} \dot{\delta a_0} \\ \vdots \\ \\ \dot{\delta a_{\mathcal{B}}} \\
\dot{\delta b_0} \\ \vdots \\ \\ \dot{\delta b_{\mathcal{B}}}
\end{array} \right\}\;.
\label{VdV}
\end{eqnarray}
and rewrite (\ref{IXY}) as 
\begin{eqnarray}
\mathcal{I}&=&\mathcal{I}_0+\vec{V}^T\mathcal{I}_2\vec{V}\nonumber\\
\mathcal{X}&=&\dot{\vec{V}^T}\mathcal{X}_2\dot{\vec{V}}\nonumber\\
\mathcal{Y}&=&\dot{\vec{V}^T}\mathcal{Y}_2\dot{\vec{V}}
\label{matIXY}
\end{eqnarray}
where $\mathcal{I}_0$ is the value of $\mathcal{I}$ for the unperturbed 
rational map as given in \cite{HMS}. 
Notice that there is no linear term in $\vec{V}$. 
This is because the unperturbed rational map minimises (\ref{I}) and 
(\ref{XYint}).

Inserting the perturbed ansatz (\ref{pertF}) and (\ref{pertRM}) into
(\ref{lag}), we get
 
\begin{eqnarray}
L=L_0+L_2
\label{pertLag}
\end{eqnarray}
where
\begin{eqnarray}
L_0&=&\frac{1}{3\pi}\int\bigg\{-2 B\sin^2f_0
    -\big(2 B\sin^2f_0+r^2\big)f_0'^2
    -\frac{\mathcal{I}_0}{r^2}\sin^4f_0\nonumber\\
   &&-\frac{2m^2r^2}{p}\big[1-\cos(pf_0)\big]\bigg\}\,dr\;,\nonumber\\
L_2&=&\frac{1}{3\pi}\int\bigg\{4r^2\dot{V}_i\mathcal{X}_{2ij}\dot{V}_j\sin^2f_0
   +2g^2 B\sin^2f_0-2g^2 B\cos^2f_0\nonumber\\
&&+4r^2\dot{V}_i\mathcal{X}_{2ij}\dot{V}_jf_0'^2\sin^2f_0-2g'^2 B\sin^2f_0
  +2g^2 B f_0'^2\sin^2f_0\nonumber\\
&&-4gg' B f_0'\sin(2f_0)-2g^2 B f_0'^2\cos^2f_0-r^2g'^2
 +2\dot{g}^2 B \sin^2f_0\nonumber\\
&&+r^2\dot{g}^2+4\dot{V}_i\mathcal{Y}_{2ij}\dot{V}_j\sin^4f_0
 +\frac{2g^2}{r^2}\mathcal{I}_0\sin^4f_0
 -\frac{6g^2}{r^2}\mathcal{I}_0(\sin^2f_0)(\cos^2f_0)\nonumber\\
&&
 -\frac{1}{r^2}V_i\mathcal{I}_{2ij}V_j\sin^4f_0-g^2m^2r^2\cos(pf_0)\bigg\}\,dr\;.
\label{pertLag12}
\end{eqnarray}

Notice that the perturbation of the radial profile and the rational map are
completely decoupled. We can thus study the radial and angular vibration modes 
separately. To do this we must compute the
Euler-Lagrange equations for $g$ and for ${\vec V}$ to linear order and 
solve the
resulting eigen value equations. 

\section{Rational Map Vibrations}
The equation for the rational map vibrations $\vec{V}$ is straightforward to 
derive from (\ref{pertLag}) and can be written as
\begin{equation}
A_{ij}\ddot{V}_j=-D_{ij}V_j\;,
\label{pertRMeq}
\end{equation}
where
\begin{eqnarray}
A_{ij} &=& \mathcal{X}_{2ij}\,\Gamma_1+\mathcal{Y}_{2ij}\,\Gamma_2\nonumber\\
D_{ij} &=& \mathcal{I}_{2ij}\,\Gamma_3
\end{eqnarray}
and
\begin{eqnarray}
\Gamma_1&=&\int(r^2\sin^2f_0+r^2f_0'^2\sin^2f_0)\,dr\nonumber\\
\Gamma_2&=&\int(\sin^4f_0)\,dr\nonumber\\
\Gamma_3&=&\int(\frac{1}{4r^2}\sin^4f_0)\,dr\,
\end{eqnarray}
are numerical parameters which can be evaluated numerically. We provide
their value for the case $m_\pi=0$ and $m_\pi=0.526$ and $p=1$ in Tables
\ref{tableG1} and \ref{tableG2}.

\begin{table}[hb]\centering
\begin{tabular}{l||l|l|l}
  &$B=1$&$B=2$&$B=4$\\
\hline
$\Gamma_1$ & 4.28869 & 7.58651 & 12.2868\\
$\Gamma_2$ & 0.872418& 0.96429 & 1.02577\\
$\Gamma_3$ & 0.370159& 0.16419 & 0.0895335\\
\end{tabular}
\caption{$\Gamma_1$, $\Gamma_2$, $\Gamma_3$, for $m_\pi=0$}
\label{tableG1}
\end{table}

\begin{table}[hb]\centering
\begin{tabular}{l||l|l|l}
  &$B=1$&$B=2$&$B=4$\\
\hline
$\Gamma_1$ & 3.04713 & 5.40344 & 8.94968\\
$\Gamma_2$ & 0.75611& 0.82419 & 0.885387\\
$\Gamma_3$ & 0.405799& 0.185219 & 0.103598\\
\end{tabular}
\caption{$\Gamma_1$, $\Gamma_2$, $\Gamma_3$, for $m_\pi=0.526$ and $p=1$}
\label{tableG2}
\end{table}

To solve (\ref{pertRMeq}), one must first compute the matrices
$\mathcal{I}_{2}, \mathcal{X}_{2}$ and $\mathcal{Y}_{2}$ defined 
in (\ref{matIXY}). Most of the entries can be shown to vanish using parity
symmetries; we have evaluated the others using both Maple and Mathematica.
 
Then we inserted $\vec{V} = \vec{V_0} \sin(\omega t)$ into (\ref{pertRMeq})
and used Maple and Mathematica to solve the resulting eigen value problem. 
We would like to point out that the eigen vectors, {\it i.e.} the vibrations, 
do not depend on the actual values on $\Gamma_1$, $\Gamma_2$ and $\Gamma_3$
(or $p$ and $m$), while the eigen values, {\it i.e.} the vibrations 
frequencies do.
The results are described separately for the cases $B=1$, 
$B=2$ and $B=4$. We will then compare our results with those of Barnes et 
al. \cite{BBT1}\cite{BBT2} obtained, in our conventions, for the value
$m_\pi=0.526$.

\subsection{B=1}
For the hedgehog solution the unperturbed ration map is simply $R=\xi$
and there are 6 perturbed rational maps:
\begin{eqnarray}
\begin{array}{lll}
R_{rot,x}(\xi) = \frac{\xi+i\,d}{1+i\,d\, \xi},&
R_{rot,y}(\xi) = \frac{\xi-d}{1+d\,\xi},&
R_{rot,z}(\xi) = \frac{\xi}{1+i\,d},
\end{array}\nonumber\\
\begin{array}{lll}
R_{btr,x}(\xi) = \frac{\xi+d}{1+d\,\xi},&
R_{btr,y}(\xi) = \frac{\xi-i\,d}{1+i\,d\,\xi},&
R_{btr,z}(\xi) = \frac{\xi}{1+d},
\end{array}
\label{RB1}
\end{eqnarray}
with 
\begin{equation}
d = \epsilon \cos(\omega t) 
\label{eqd}
\end{equation}
where $\epsilon$ and $\omega$ are respectively the (small) amplitude 
and the frequency of vibration.

$R_{rot,x}$, $R_{rot,y}$ and $R_{rot,z}$ are zero modes and correspond respectively to
a rotation around the $x$, $y$ and $z$ axis.
$R_{btr,x}$, $R_{btr,y}$ and $R_{btr,z}$ have the same eigen value 
$\omega_{btr} = 0.7575$ for $m_\pi=0$ and $\omega_{btr} = 0.9239$ for $m_\pi=0.526$.
They form a 3 dimensional eigen space as they are conjugated to each other 
through a $90^{\circ}$ rotation. This eigen subspace corresponds to the
broken translational invariance of the solutions: while the Skyrme Lagrangian
(\ref{lag}) is invariant under translation, the rational map ansatz breaks
that symmetry by pinning the center of the solution. If one performs the 
translation $x \rightarrow x+x_0$, into the hedgehog ansatz, the
ansatz is broken, as the radial profile becomes a function of the polar 
angles and the rational map becomes a function of $r$. If $x_0$ is small,
one can expand the translated expression, keeping only the linear terms in 
$x_0$. The rational map then becomes 
$R(\xi) = \frac{\xi+\epsilon/r}{1+\epsilon \xi /r}$.
As the energy density of the Skyrmion is concentrated around a shell of 
radius $r_0$, we can take 
$R(\xi) = \frac{\xi+\epsilon/r_0}{1+ \epsilon \xi /r_0}$ as an 
approximation of the translated rational map, recovering $R_{btr,x}(\xi)$. As we 
had to make several approximations to derive that expression of the perturbed 
rational map, instead of being a zero mode, it has a non zero vibration 
frequency that does not have any physical meaning.

We can thus conclude that, as expected, the rational map of the hedgehog 
solution does not have any genuine non-zero vibration mode.

\subsection{B=2} 
The unperturbed rational map for the case $B=2$ is given by $R=\xi^2$ and
there are 10 perturbed rational maps.

First we have the 3 rotational zero modes
\begin{eqnarray}
R_{rot,x}(\xi) = \frac{\xi^2+i\,\epsilon\,\xi}{1+i\,\epsilon\, \xi},\qquad\qquad
R_{rot,y}(\xi) = \frac{\xi^2-\epsilon\,\xi}{1+\epsilon\, \xi},\qquad\qquad
R_{rot,z}(\xi) = \frac{\xi^2}{1+i \,\epsilon}.
\label{RB2rot}
\end{eqnarray}

Then there are 2 iso-rotation zero modes
\begin{eqnarray}
R_{iso,x}(\xi) = \frac{\xi^2+i\,\epsilon}{1+i\,\epsilon\, \xi^2},\qquad\qquad
R_{iso,y}(\xi) = \frac{\xi^2-\epsilon}{1+\epsilon\,\xi^2}.
\label{RB2iso}
\end{eqnarray}
There are only 2 such modes because the iso-rotation around the $z$ axis
coincides with the proper rotation around the $z$ axis.

The lowest vibration modes are
\begin{eqnarray}
R_{btr,x}(\xi) = \frac{\xi^2+\epsilon\, \xi}{1+\epsilon\, \xi},\qquad\qquad
R_{btr,y}(\xi) = \frac{\xi^2-i\,\epsilon\,\xi}{1+i\,\epsilon\, \xi},\qquad\qquad
R_{btr,z}(\xi) = \frac{\xi^2}{1+\epsilon},
\label{RB2tr}
\end{eqnarray}
but they correspond to the broken translation modes.
Their eigen values $\omega_{btr}$  are respectively $0.69$, $0.69$ and 
$0.4766$ for $m_\pi=0$ and $0.847$ $0.847$ and $0.584$  for $m_\pi=0.526$,
but they have no physical meaning.

\begin{figure}[!htp]
\begin{minipage}{1.0\textwidth}\hspace*{0.5 cm}
\begin{minipage}[t]{0.38\textwidth}
\includegraphics[width=1.25\textwidth]{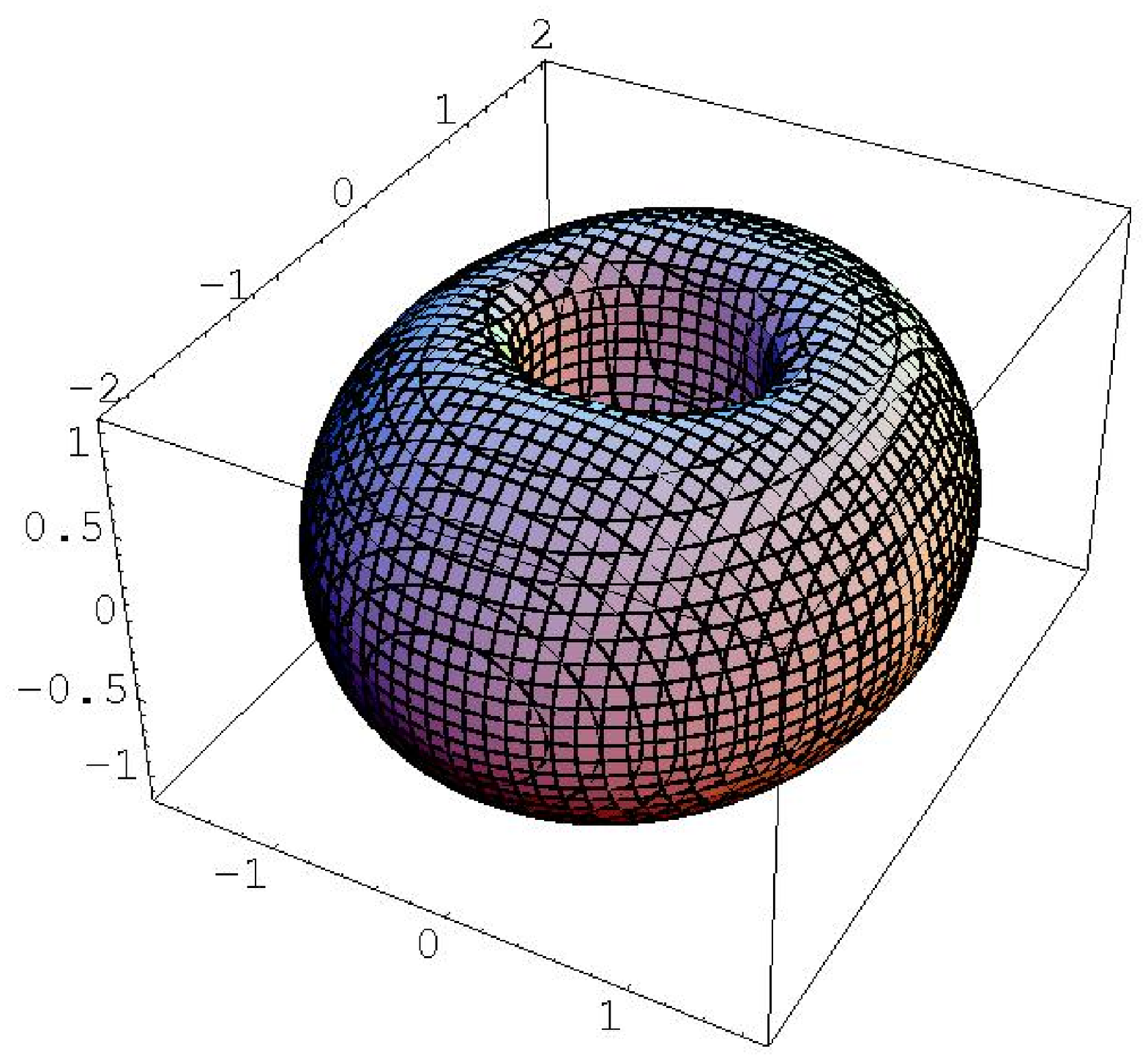}
\end{minipage}
\hspace*{1.5cm}
\begin{minipage}[t]{0.38\textwidth}
\includegraphics[width=1.25\textwidth]{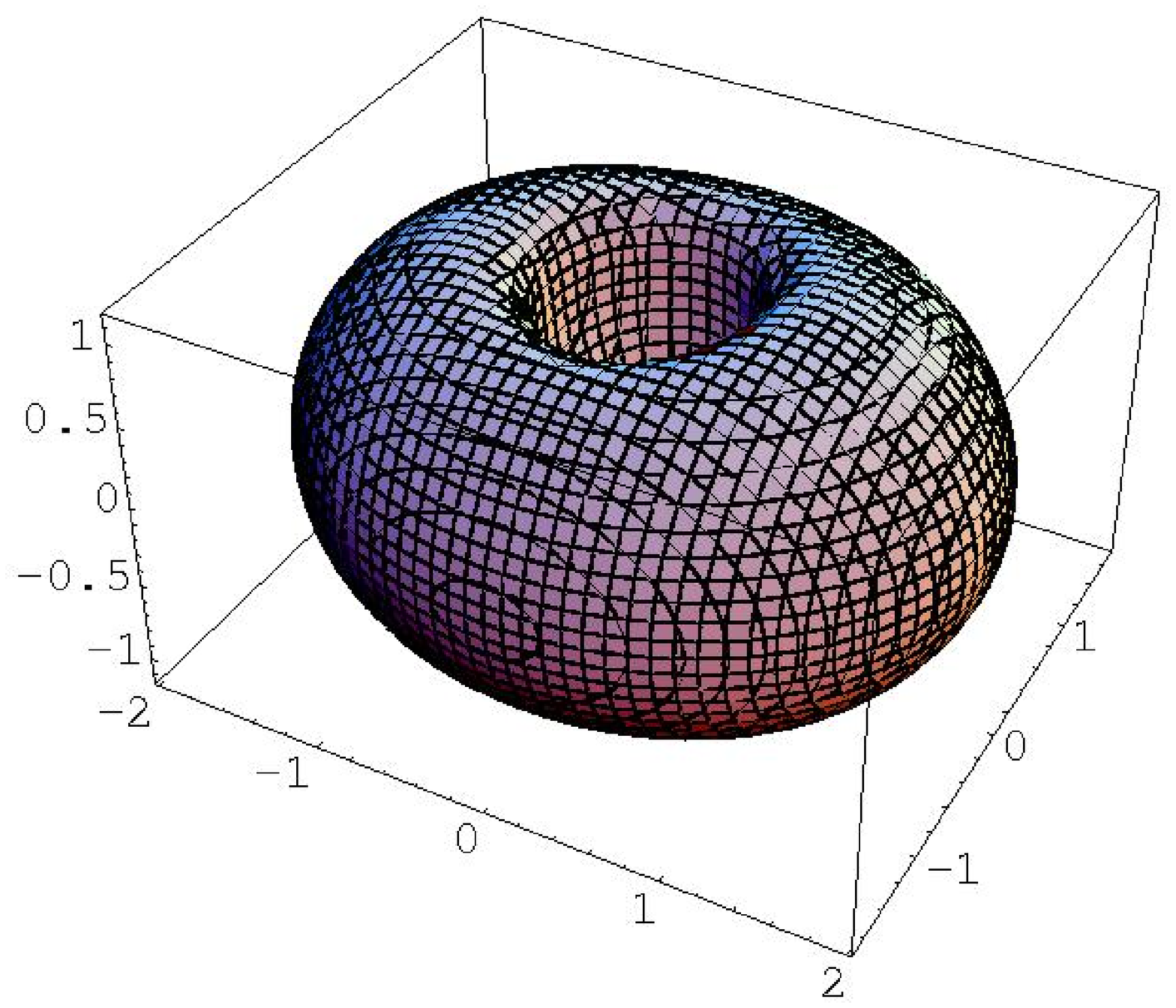}
\end{minipage}
\end{minipage}
\caption{$B=2$, $R_{s1}(\xi)$:  (Left) $\epsilon >0$; (Right)
$\epsilon <0$.}\label{fig1_B2Rs}
\end{figure}

We are thus left with a 2 parameter class of genuine vibration modes:
\begin{eqnarray}
R_{s1}(\xi) = \frac{\xi^2+\epsilon}{1+\epsilon\, \xi^2},\qquad\qquad
R_{s2}(\xi) = \frac{\xi^2-i\,\epsilon\,}{1+i\,\epsilon\,\xi^2}.
\label{RB2scat}
\end{eqnarray}
These two perturbed rational maps correspond to lateral squeezing and 
stretching 
of the torus alternating between elongations along the $x$ and $y$ axis as
shown in Figure \ref{fig1_B2Rs}. This mode is sometimes referred to as the 
scattering mode as it corresponds to two Skyrmions colliding with each other.
$R_{s2}$ corresponds to the same deformation but rotated by $45^{\circ}$. By 
taking 
a linear combination of  $R_{s1}$ and $R_{s2}$ the torus can be made to wobble 
along any axis in the $x-y$ plane. The eigen value for these 2 modes
is $\omega_{s} = 0.9909$ for $m_\pi=0$ and $\omega_{s} = 1.2255$ for 
$m_\pi=0.526$.

\subsection{B=4} 
The unperturbed rational map for the case $B=4$ is given by $R=P_0/Q_0$ 
where $P_0=\xi^4+2\sqrt 3\;i\;\xi^2+1$ and $Q_0=\xi^4-2\sqrt3\;i\;\xi^2+1$.
There are 18 perturbed rational maps, including 6 zero modes:
\begin{eqnarray}
R_{rot,x}(\xi)&=&
  \frac{P_0+(2i\,\theta_{x}-2\sqrt3\;\theta_{x})\xi^3
       +(2i\,\theta_{x}-2\sqrt3\;\theta_{x})\xi}
     {Q_0+(2i\,\theta_{x}+2\sqrt3\;\theta_{x})\xi^3
      +(2i\,\theta_{x}+2\sqrt3\;\theta_{x})\xi},
\nonumber\\
R_{rot,y}(\xi)&=&
\frac{P_0+(2\sqrt3\;i\,\theta_{y}-2\theta_{y})\xi^3
      +(-2\sqrt3\;i\,\theta_{y}+2\theta_{y})\xi}
     {Q_0+(-2\sqrt3\;i\,\theta_{y}-2\theta_{y})\xi^3
      +(2\sqrt3\;i\,\theta_{y}+2\theta_{y})\xi},
\nonumber\\
R_{rot,z}(\xi)&=&\frac{P_0+4\sqrt3\;\theta_{z}\;\xi^2-4\;i\,\theta_{z}}
                  {Q_0-4\sqrt3\;\theta_{z}\;\xi^2-4\;i\,\theta_{z}},
\nonumber\\
R_{iso,x}(\xi)&=&\frac{P_0+2\sqrt{3}\theta_1\,\xi^2}
             {Q_0-2\sqrt{3}\theta_1\,\xi^2},
\nonumber\\
R_{iso,y}(\xi)&=&\frac{P_0-2\sqrt{3}\,i\theta_2\,\xi^2}
                  {Q_0-\theta_2\xi^4-\theta_2},
\nonumber\\
R_{iso,z}(\xi)&=&\frac{P_0\cdot(1+i\theta_3)}{Q_0},
\label{RB4zm}
\end{eqnarray}
which correspond to the 3 rotations $R_{rot,x}$, $R_{rot,y}$ and $R_{rot,z}$ and 3 
iso-rotations $R_{iso,x}$, $R_{iso,y}$ and $R_{iso,z}$.  

The lowest vibration modes are the broken translation
\begin{eqnarray}
R_{btr,x}(\xi)&=&
  \frac{P_0-(\frac{1}{2}+\frac{\sqrt{3}\,i}{2})\,\epsilon\,\xi^3
   -(\frac{1}{2}+\frac{\sqrt{3}\,i}{2})\,\epsilon\,\xi}
    {Q_0-(\frac{1}{2}-\frac{\sqrt{3}\,i}{2})\epsilon\,\xi^3
     -(\frac{1}{2}-\frac{\sqrt{3}\,i}{2})\,\epsilon\,\xi},
\nonumber\\
R_{btr,y}(\xi)&=&
   \frac{P_0-(\frac{\sqrt{3}}{2}+\frac{i}{2})\,\epsilon\,\xi^3
   +(\frac{\sqrt{3}}{2}+\frac{i}{2})\,\epsilon\,\xi}
    {Q_0+(\frac{\sqrt{3}}{2}-\frac{i}{2})\epsilon\,\xi^3
     -(\frac{\sqrt{3}}{2}-\frac{i}{2})\,\epsilon\,\xi},
\nonumber\\
R_{btr,z}(\xi)&=&\frac{P_0+\sqrt{3}\,i\,\epsilon\,\xi^2+\epsilon\,}
    {Q_0-\sqrt{3}\,i\,\epsilon\,\xi^2+\epsilon}
\end{eqnarray}
and their eigen values are $\omega_{tr} = 0.4586$ for $m_\pi=0$ 
and $\omega_{tr} = 0.5625$ for $m_\pi=0.526$.

\begin{figure}[!htp]
\begin{minipage}{1.0\textwidth}\hspace*{0.5 cm}
\begin{minipage}[t]{0.38\textwidth}
\includegraphics[width=1.25\textwidth]{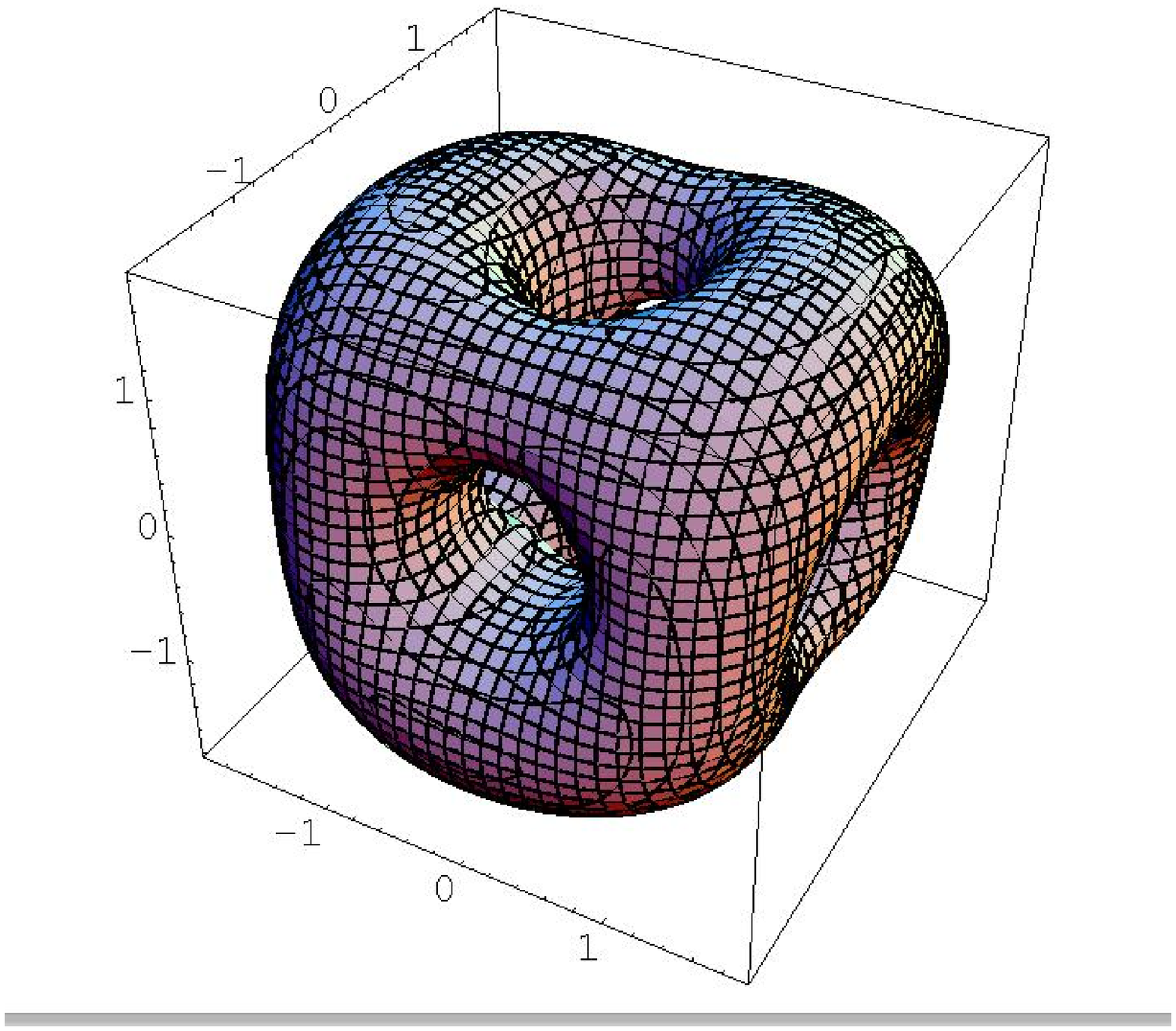}
\end{minipage}
\hspace*{1.5cm}
\begin{minipage}[t]{0.38\textwidth}
\includegraphics[width=1.25\textwidth]{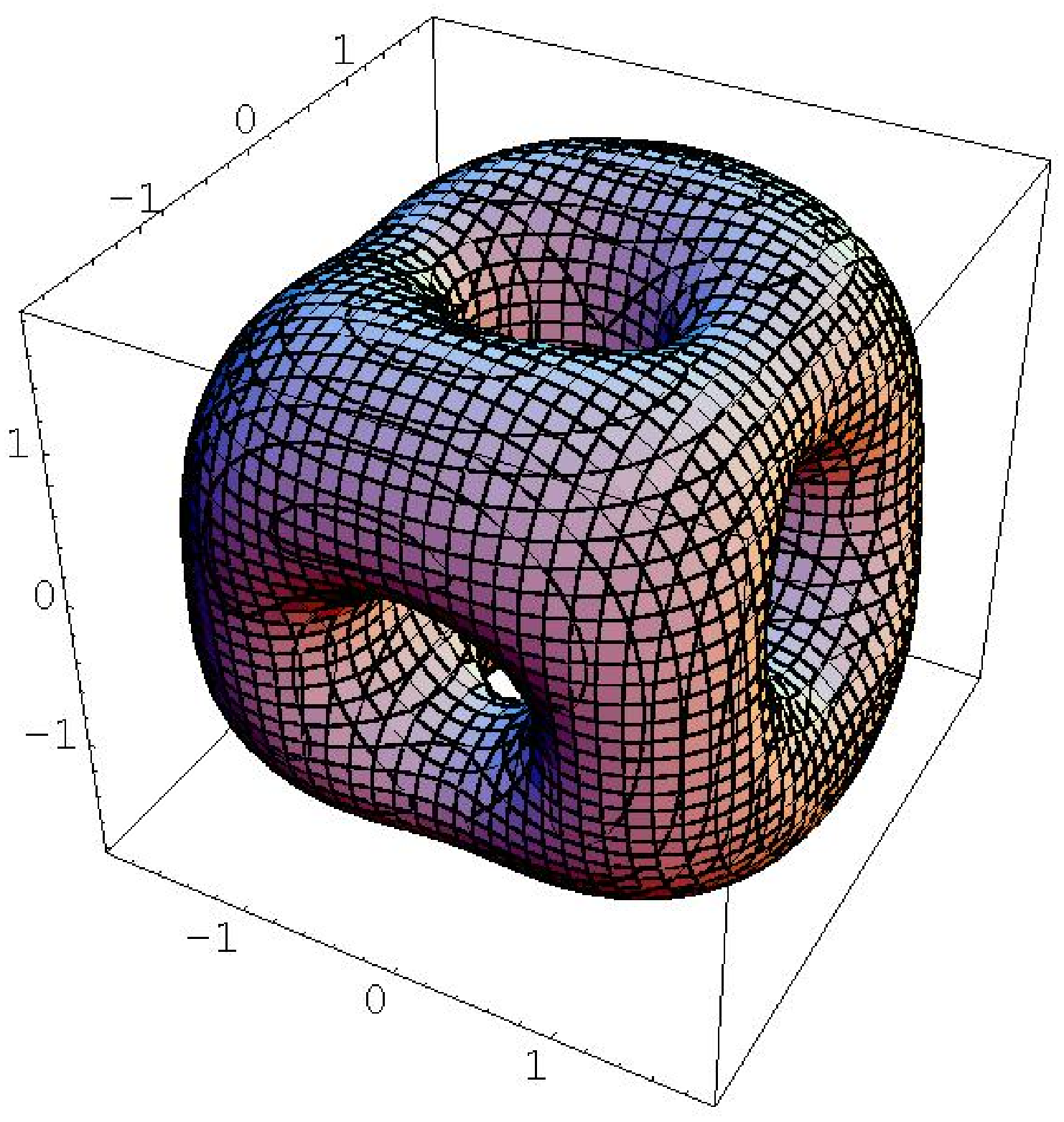}
\end{minipage}
\end{minipage}
\caption{$B=4$, $R_{th,z}(\xi)$:  (Left) $\epsilon >0$; (Right)
$\epsilon <0$.}\label{fig2_B4Rth}
\end{figure}

The first genuine vibration mode is given by
\begin{eqnarray}
R_{tb,x}(\xi)&=&\frac{P_0-(\frac{\sqrt{3}}{2}-\frac{i}{2})\,\epsilon\,\xi^3
   -(\frac{\sqrt{3}}{2}-\frac{i}{2})\,\epsilon\,\xi}
    {Q_0-(\frac{\sqrt{3}}{2}+\frac{i}{2})\epsilon\,\xi^3
    -(\frac{\sqrt{3}}{2}+\frac{i}{2})\,\epsilon\,\xi},
\nonumber\\
R_{tb,y}(\xi)&=&
  \frac{P_0-(\frac{1}{2}-\frac{\sqrt{3}\,i}{2})\,\epsilon\,\xi^3
   +(\frac{1}{2}-\frac{\sqrt{3}\,i}{2})\,\epsilon\,\xi}
   {Q_0+(\frac{1}{2}+\frac{\sqrt{3}\,i}{2})\epsilon\,\xi^3
   -(\frac{1}{2}+\frac{\sqrt{3}\,i}{2})\,\epsilon\,\xi},
\nonumber\\
R_{tb,z}(\xi)&=&\frac{P_0+\sqrt{3}\,\epsilon\,\xi^2-i\,\epsilon}
  {Q_0-i\,\epsilon\,\xi^4-\sqrt{3}\,\epsilon\,\xi^2}
\end{eqnarray}
and its eigen value is $\omega_{tb} = 0.6093$ for $m_\pi=0$ and 
$\omega_{tb} =0.7508$ for $m_\pi=0.526$.
To picture it one, has to think of the cube as two tori stacked on top of 
each other. Then each torus oscillates like the scattering mode of the $B=2$ 
solution but in phase opposition. This results in a cube which is somewhat 
squeezed into a toroidal shape as shown in figure \ref{fig2_B4Rth}. 
This can also be thought of as the scattering of four Skyrmions in a 
tetrahedral configuration.

\begin{figure}[!htp]
\begin{minipage}{1.0\textwidth}\hspace*{0.5 cm}
\begin{minipage}[t]{0.38\textwidth}
\includegraphics[width=1.25\textwidth]{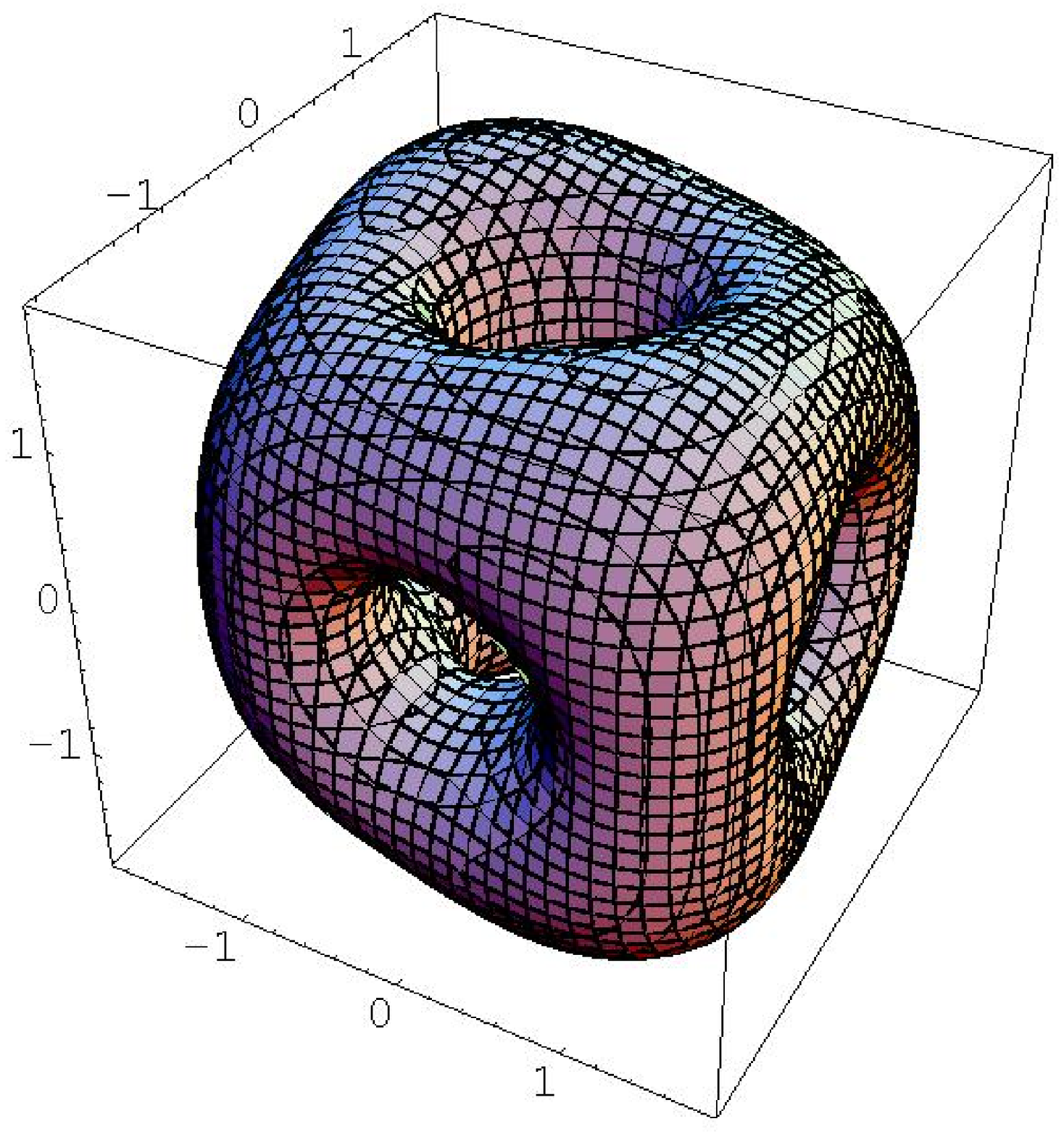}
\end{minipage}
\hspace*{1.5cm}
\begin{minipage}[t]{0.38\textwidth}
\includegraphics[width=1.25\textwidth]{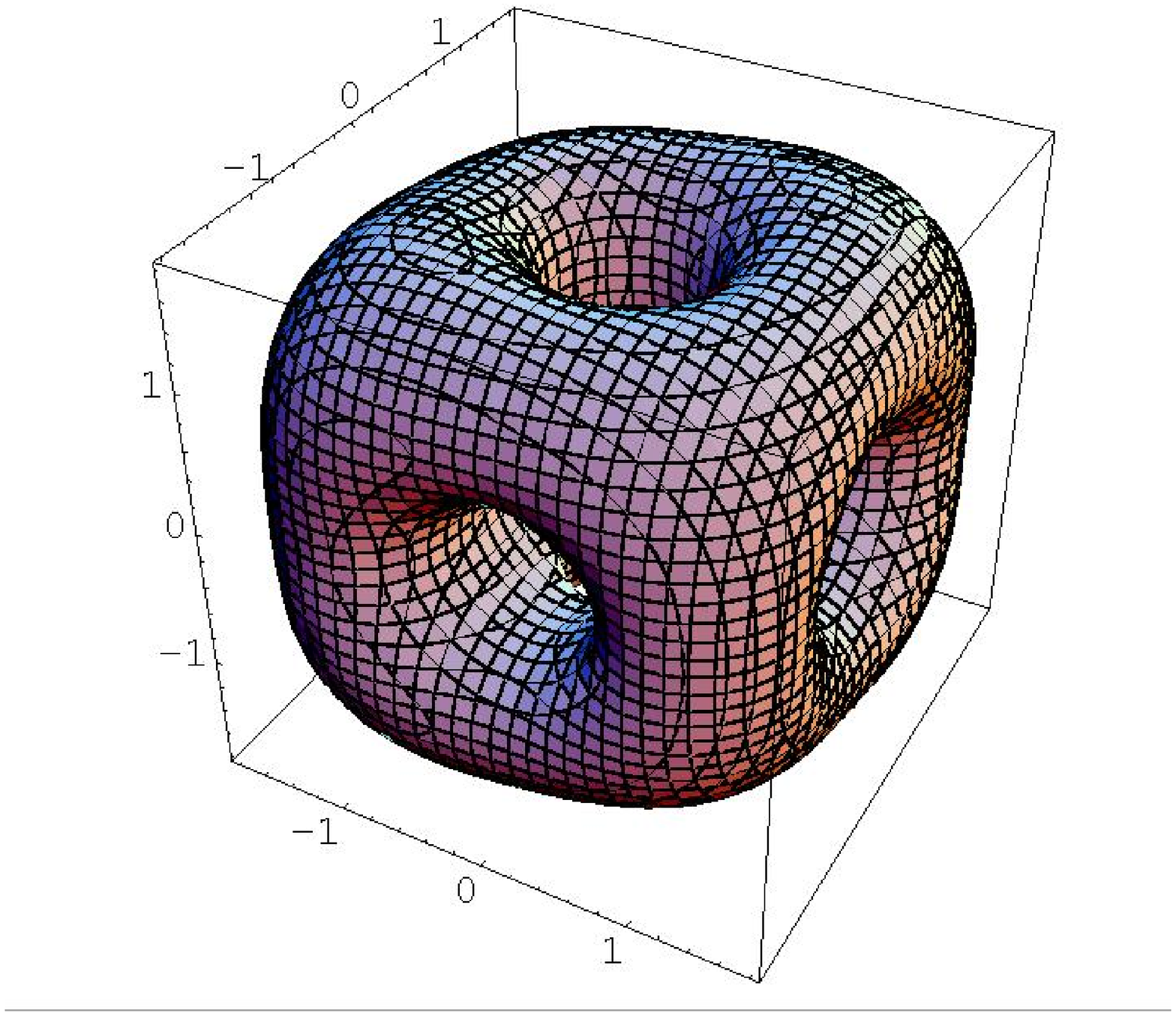}
\end{minipage}
\end{minipage}
\caption{$B=4$, $R_{rhomb,z}(\xi)$:  (Left) $\epsilon >0$; (Right)
$\epsilon <0$.}\label{fig3_B4Rrh}
\end{figure}

The last triplet mode corresponds to the pinching and stretching of two 
opposite edges of the cube so that two of the opposite faces of the cube are 
deformed into a rhombus.
\begin{eqnarray}
R_{rhomb,x}(\xi)&=&
   \frac{P_0+(\frac{1}{2}+\frac{\sqrt{3}\,i}{2})\,\epsilon\,\xi^3
    +(\frac{1}{2}+\frac{\sqrt{3}\,i}{2})\,\epsilon\,\xi}
   {Q_0-(\frac{1}{2}-\frac{\sqrt{3}\,i}{2})\epsilon\,\xi^3
    -(\frac{1}{2}-\frac{\sqrt{3}\,i}{2})\,\epsilon\,\xi},
\nonumber\\
R_{rhomb,y}(\xi)&=&
  \frac{P_0+(\frac{\sqrt{3}}{2}+\frac{i}{2})\,\epsilon\,\xi^3
        -(\frac{\sqrt{3}}{2}+\frac{i}{2})\,\epsilon\,\xi}
       {Q_0+(\frac{\sqrt{3}}{2}-\frac{i}{2})\epsilon\,\xi^3
        -(\frac{\sqrt{3}}{2}-\frac{i}{2})\,\epsilon\,\xi},
\nonumber\\
R_{rhomb,z}(\xi)&=&\frac{P_0-\sqrt{3}\,i\,\epsilon\,\xi^2-\epsilon}
      {Q_0-\epsilon\,\xi^4+\sqrt{3}\,i\,\epsilon\,\xi^2}.
\end{eqnarray}
The energy density plots are presented in figure \ref{fig3_B4Rrh}. 
The eigen value of the rhombus mode is $\omega_{rhomb} = 0.7395$
 for $m_\pi=0$ and $\omega_{rhomb} = 0.9113$ for $m_\pi=0.526$.

\begin{figure}[!htp]
\begin{minipage}{1.0\textwidth}\hspace*{0.5 cm}
\begin{minipage}[t]{0.38\textwidth}
\includegraphics[width=1.25\textwidth]{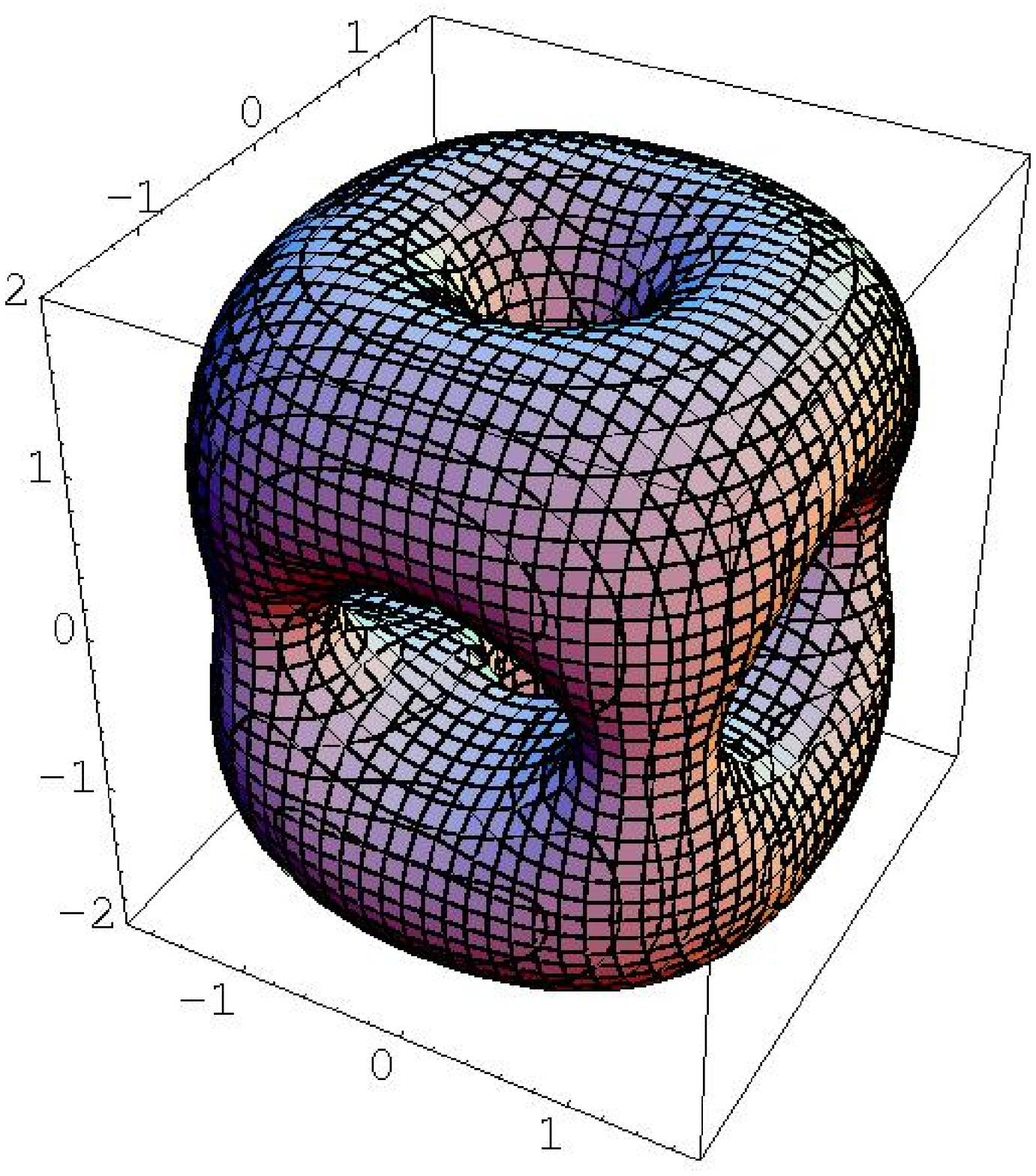}
\end{minipage}
\hspace*{1.5cm}
\begin{minipage}[t]{0.38\textwidth}
\includegraphics[width=1.25\textwidth]{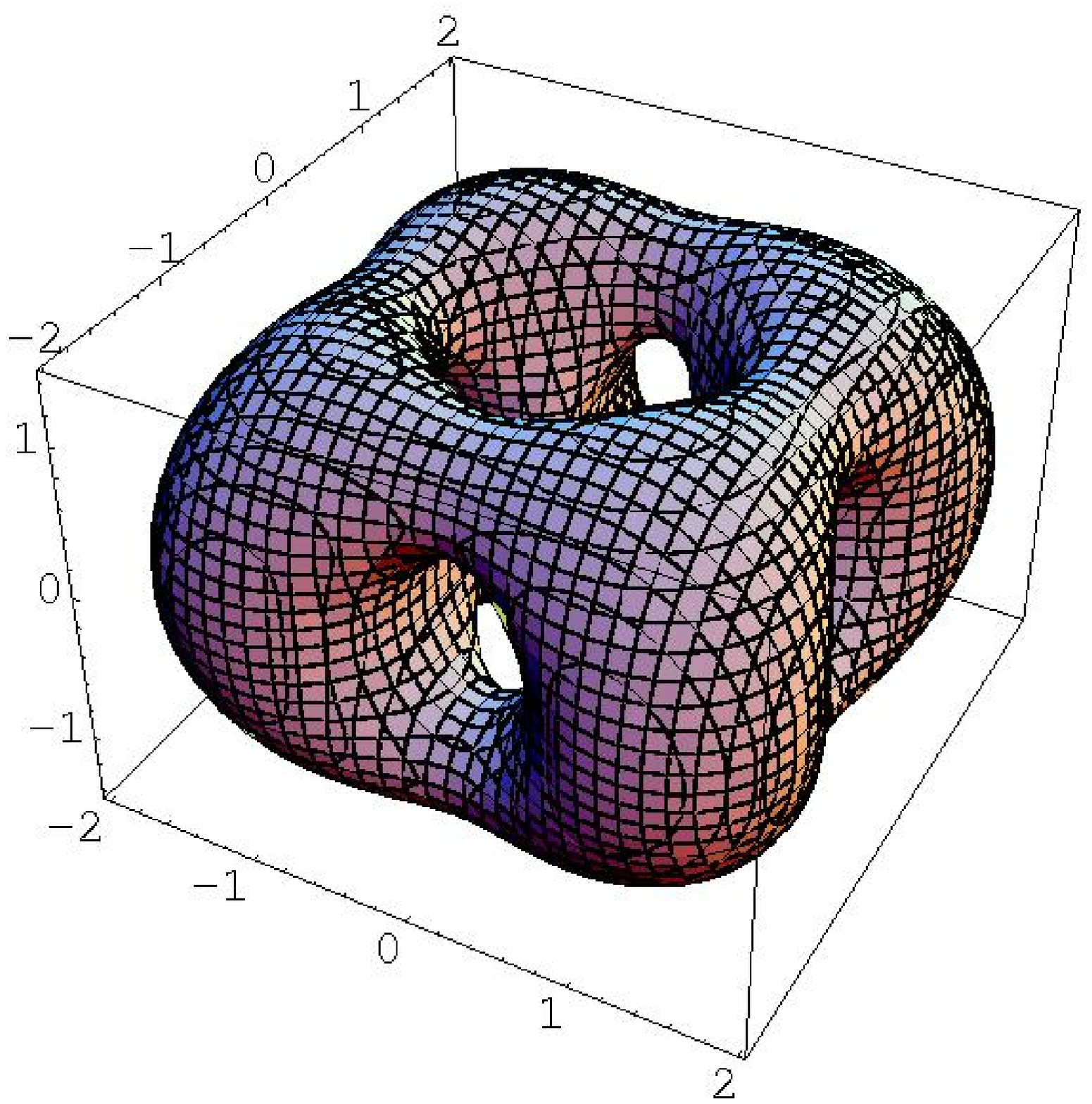}
\end{minipage}
\end{minipage}
\caption{$B=4$, $R_{scat2}(\xi)$:  (Left) $\epsilon >0$; (Right)
$\epsilon <0$.}\label{fig4_B4Rsc}
\end{figure}

The doublet mode 
\begin{eqnarray}
R_{scat1}(\xi)&=&\displaystyle\frac{P_0-2\sqrt{3}\,\epsilon\,\xi^2}
    {Q_0+i\,\epsilon\,\xi^4+i\,\epsilon},
\nonumber\\
R_{scat2}(\xi)&=&\displaystyle\frac{P_0+i\,\epsilon\,\xi^2}
    {Q_0-i\,\epsilon\,\xi^2},
\end{eqnarray}
corresponds to a deformation where the cube is alternatively stretched and then 
flattened along one of the symmetry axis going through the center of the cube's 
faces. A combination of a stretch along two of the (perpendicular) axis is 
equivalent to a stretch along the third axis. This corresponds to the 
scattering 
mode of two tori along their axis of symmetry as shown in figure 
\ref{fig4_B4Rsc}.
The eigen value for this class of modes is $\omega_{str} = 0.818$ 
for $m_\pi=0$ and $\omega_{str} =1.0064$ for $m_\pi=0.526$.

\begin{figure}[!htp]
\begin{minipage}{1.0\textwidth}\hspace*{0.5 cm}
\begin{minipage}[t]{0.38\textwidth}
\includegraphics[width=1.25\textwidth]{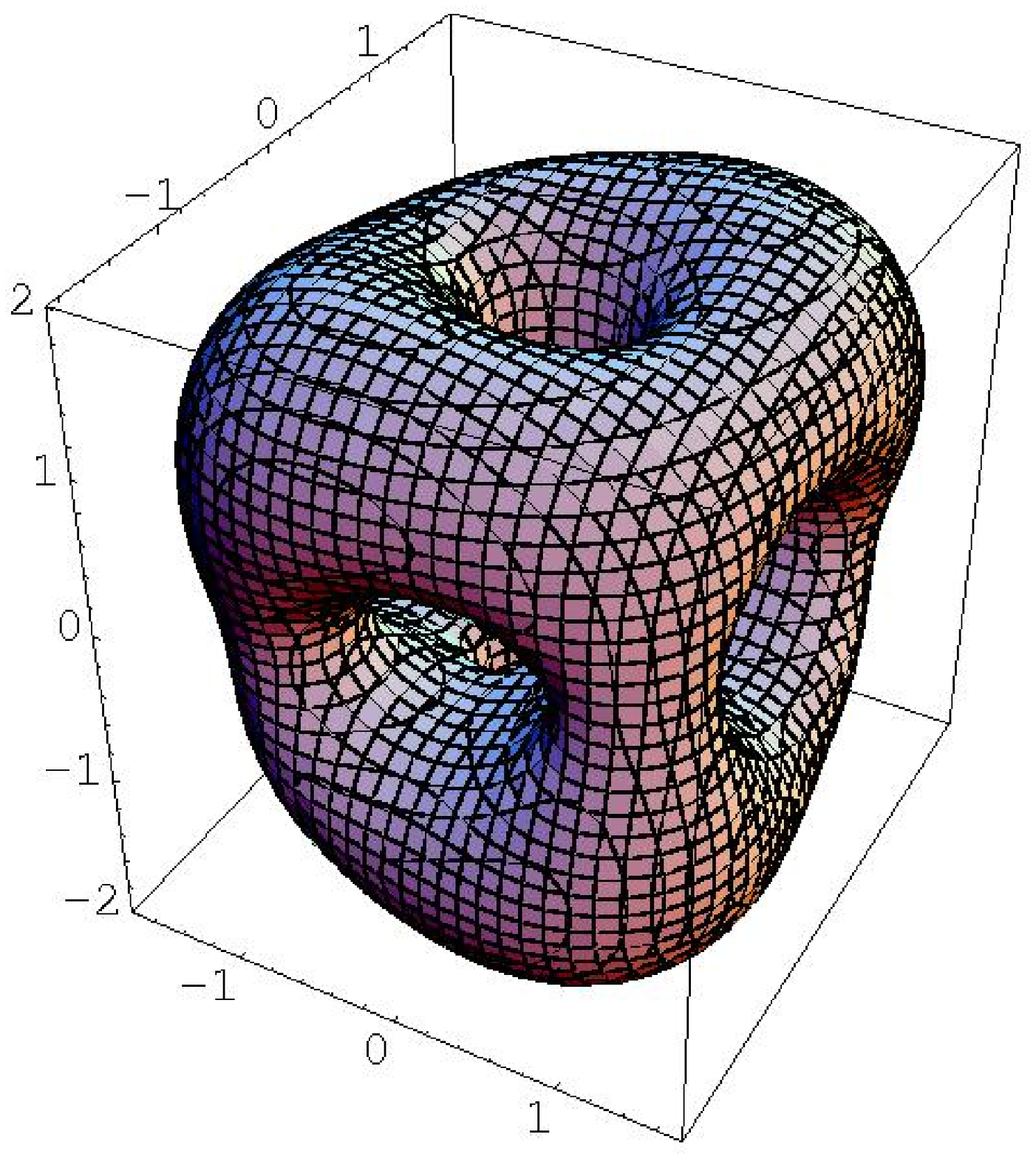}
\end{minipage}
\hspace*{1.5cm}
\begin{minipage}[t]{0.38\textwidth}
\includegraphics[width=1.25\textwidth]{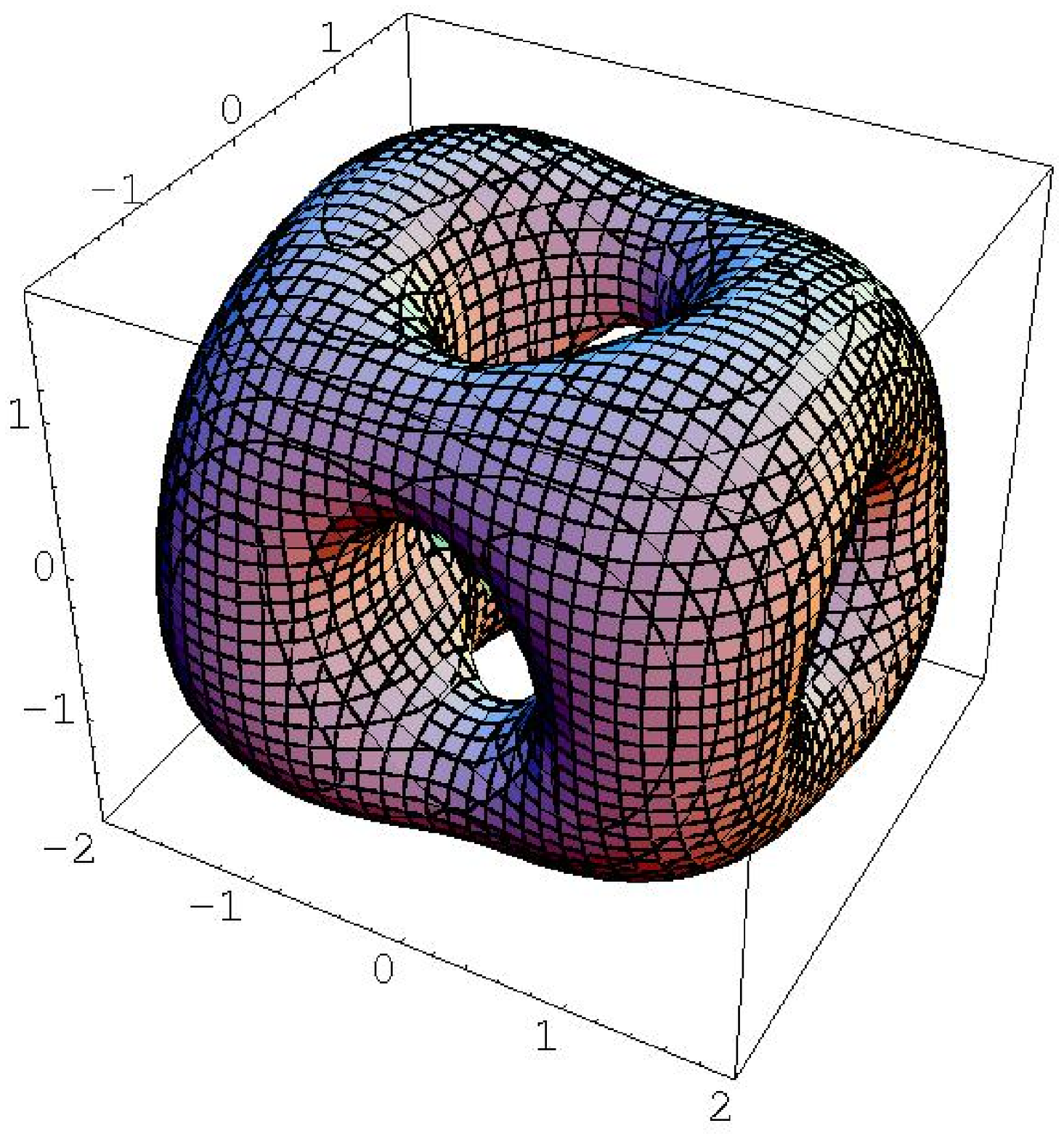}
\end{minipage}
\end{minipage}
\caption{$B=4$, $R_{tet}(\xi)$:  (Left) $\epsilon >0$; (Right)
$\epsilon <0$.}\label{fig5_B4Rtet}
\end{figure}

The singlet mode 
\begin{eqnarray}
R_{tet}&=&\frac{P_0}{Q_0(1-\epsilon)}
\end{eqnarray}
corresponds to a tetrahedral deformation of the cube as illustrated in Figure 
\ref{fig5_B4Rtet}. Its eigen value is $\omega_{tet} = 1.1355$ 
for $m_\pi=0$ and $\omega_{tet} = 1.4026$ for $m_\pi=0.526$.

\section{Radial Vibrations}
The Euler-Lagrange equation for the radial perturbation $g$ can be easily
derived from (\ref{pertLag}):
\begin{eqnarray}
&&(\ddot{g}-g'')\,\Bigl(2\mathcal{B}\sin^2f_0+r^2\Bigr)
-\Bigl(2r+2\mathcal{B}f'_0\sin(2f_0)\Bigl)\,g'
-\Bigl(-2\mathcal{B}\cos(2f_0)+2\mathcal{B}f'^2_0\cos(2f_0)\nonumber\\
 &&+2\mathcal{B}f''_0\sin(2f_0) 
 -\frac{\mathcal{I}_0}{r^2}\Big[6(\sin^2f_0)\,
  (\cos^2f_0)-2\sin^4f_0\Big]-m^2r^2\cos(pf_0)
  \Bigr)\,g\;= 0,
\label{pertFeq}
\end{eqnarray}
where $g'=\frac{\partial}{\partial r}$ and 
$\dot{g}=\frac{\partial}{\partial t}$.
Taking a perturbation of the form
\begin{equation}
g(r,t) = g_0(r)\, \sin(\omega t)
\end{equation}
and inserting it into (\ref{pertFeq}) leads to a Sturm-Liouville
equation where $\omega^2$ is the eigen value. 

Notice that when $\omega > m$, any perturbation is radiated away and there
are no genuine vibration modes. When the pion
mass $m_\pi$ is too small, there are thus no genuine radial vibration modes.
On the other hand, the Skyrmion still has so-called 
pseudo-vibration modes, as studied by Bizon et al.\cite{BCR}, where the 
excitation energy is radiated away relatively slowly. In a quantum theory 
these modes would correspond to resonances.

\begin{figure}[!htp]
\begin{center}
\includegraphics[width=0.7\textwidth]{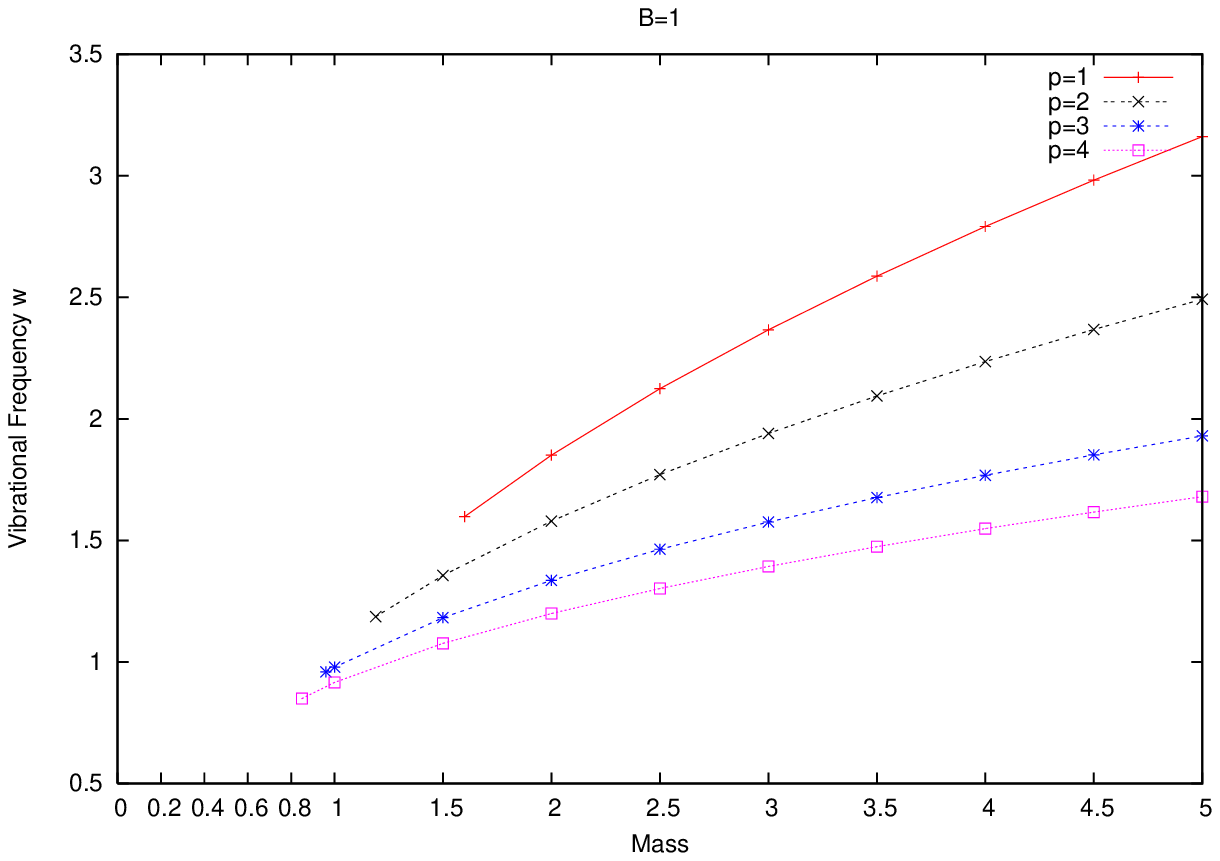}
\includegraphics[width=0.7\textwidth]{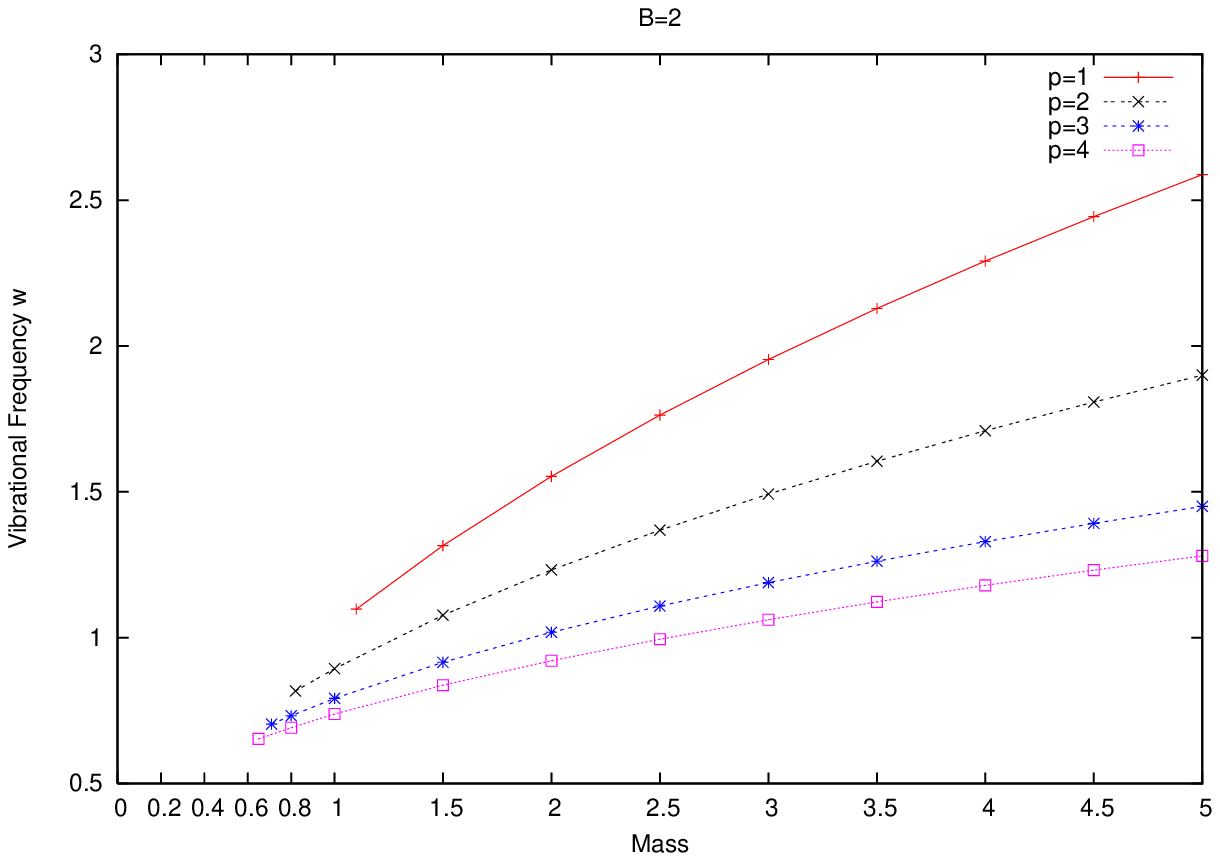}
\includegraphics[width=0.7\textwidth]{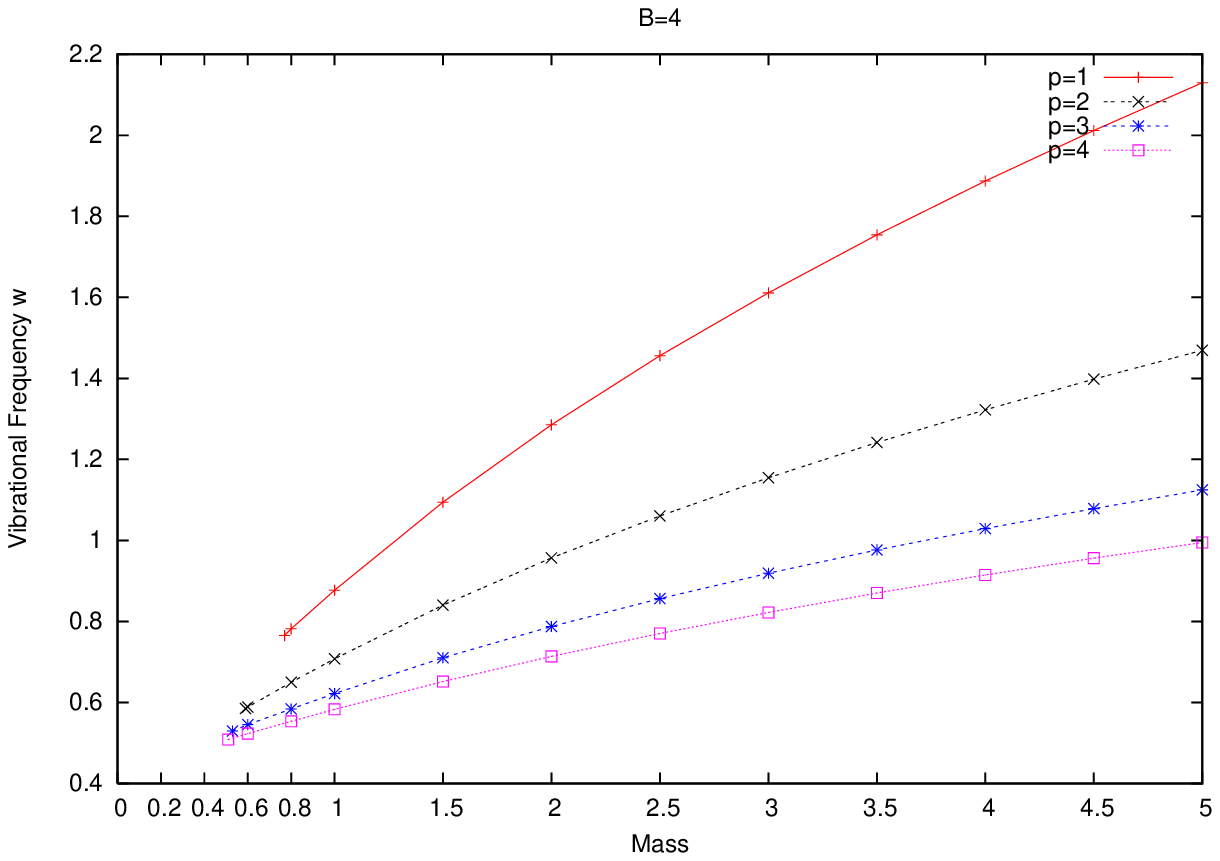}
\caption{Radial Vibration frequency for 
$\mathcal{B}$=1$, \mathcal{B}$=2, $\mathcal{B}$=4.}
\label{fig6}
\end{center}
\end{figure}

We have solved (\ref{pertFeq}) numerically for several baryon numbers
and several values of the pion mass and the parameter $p$. Our results are 
summarised in figure \ref{fig6} where we present the vibration frequency 
$\omega$ as a 
function of the pion mass for different baryon number values.

On table \ref{tableG3} we present the critical value of the pion mass below 
which 
there is no genuine vibration mode. Notice that the case $B=1$ is the exact 
hedgehog solution and corresponds to a single nucleon. The critical value of
the mass for that case, can thus be interpreted as an upper bound for the 
pion mass when it is taken as an adjustable parameters \cite{BKS}\cite{BMS}, 
as the proton has no stable excited state.

\begin{table}\centering
\begin{tabular}{l|llll}
\hline
$B$ $\setminus$ p&1&2&3&4\\
\hline
\hline
1&1.60&1.19&0.96&0.85\\
\hline
2&1.10&0.82&0.71&0.65\\
\hline
4&0.77&0.59&0.53&0.51\\
\hline
6&0.60&0.47&0.44&0.43\\
\hline
7&0.56&0.44&0.42&0.41\\
\hline
9&0.47&0.38&0.36&0.36\\
\hline
17&0.33&0.28&0.27&0.27
\end{tabular}
\caption{Critical Mass as a function of $B$ and $p$.}
\label{tableG3}
\end{table}

\section{Comparison}
Several years ago, Barnes et al\cite{BBT1}\cite{BBT2} computed the 
vibration modes 
of the $B=2$ and $B=4$ Skyrmion solutions numerically. 
In their work, they only considered the standard mass term, $p=1$ 
and the pion mass that they used, in our parametrisation is given 
by $m_\pi = 0.526$.
By comparing their
results with ours, we will be able to asses the quality of the rational map 
for the estimation of the vibration modes. 

The method used by Barnes et al. consists in solving numerically small 
fluctuations of the Skyrme field around a static solution 
obtained numerically too. To solve the Skyrme equation numerically, one has 
to invert a matrix that is field-dependant. To improve the efficiency of 
their code, and motivated by the fact that they only studied small vibrations,
Barnes et al. set that matrix to the one obtained from the stationary field. 
This effectively pinned the position and iso-orientation 
of the Skyrmion, thus breaking the translation and iso-rotation zero modes.
On the other hand, the rotations symmetry was preserved, apart from the 
symmetry breaking introduced by the square lattice and the periodic boundary 
conditions.

We compare our results with those of \cite{BBT1} and \cite{BBT2} in 
Tables \ref{table4} and \ref{table5}. One can identify the various vibration 
modes by comparing the dimension of the subspace they span, {\it i.e.} the 
dimension of the representation they belong to, as well as
their symmetries and the deformation induced in energy density plots. 
Figures 2 in \cite{BBT1}
match exactly those we have obtained for the vibrational modes of the rational
map of $B=4$. Figure 3 in \cite{BBT2}, once translated to what is plotted in 
figure \ref{fig1_B2Rs}, also matches our observation for $B=2$.

For $B=2$, we had to read the values of $\omega$ from figure 1 in \cite{BBT2}.
There is only one genuine vibration mode, the so-called scattering mode.
The frequency obtained from the rational map for that mode is much higher 
than the one obtained numerically, but the frequency obtained from 
Barnes et al. indicates that it is a genuine vibration mode. 

The mode described as a dipole breathing motion in \cite{BBT2}, turns out to 
be a broken translation. 
Notice also that the breathing mode is a pseudo vibration mode for both the 
numerical and the rational map results. The radiative modes observed in 
\cite{BBT2} are the results of the numerical method used and are excluded 
from the rational map ansatz, except from the mode with symmetry $A_{1u}$ which 
happens to be a broken translation mode.

For $B=4$, the modes that we obtained match those obtained in \cite{BBT1}.
The mode with $\omega_{num}=0.655$ in \cite{BBT1} turns out to be a broken 
translation mode rather than a proper vibration mode. This is supported by 
comparing the energy densities for that 
mode, not presented in this paper, and the one given in \cite{BBT1}, as well 
as the symmetry of that mode. Apart from that, the rational map ansatz
predicts 5 non-null vibration modes out of 6 (the diagonal mode, with 
$\omega_{num}=0.738$, is missing). Symmetry considerations suggest that the 
radiative mode with $\omega_{num}=0.587$ might be a pseudo-iso-rotational mode, 
but it is difficult to be conclusive as one would need
to know more about this numerical mode to make a definite statement.
If the vibration modes obtained from the rational map ansatz match the one 
obtained numerically rather well, the predictions for the frequencies of 
these modes are rather poor and are not even in the correct order. 
Overall,
the rational map ansatz appears to be stiffer than real Skyrmion solutions.

\begin{sidewaystable}
\begin{center}
\begin{threeparttable}\fontsize{9}{12pt}\selectfont
\begin{tabular}{cclc|cclc}
\toprule
&&&&&\\
&&Numerical\cite{BBT2}&&&Ansatz&&\\
&&&&&\\
\midrule
&&&&&\\
$\omega$&Degeneracy&Description&Symmetry&$\omega$&Degeneracy&Description&
               Symmetry\\
&&&&&\\
0.03&2&broken rotational modes&$E_{1g}$&0&3&
               $x$, $y$ rotational modes $(R_{rot,x}, R_{rot,y})$&$E_{1g}+A_{2g}$\\
&&(around $x$ and $y$ axes)&&&&and $z$ (iso)rotational mode $(R_{rot,z})$&\\
0.31&2&2 Skyrmions scattering mode&$E_{2g}$&1.225536&2&
               2 Skyrmions scattering mode $(R_{s1}, R_{s2})$&$E_{2g}$\\
0.49&1&radiative mode&$A_{1u}$&&&&\\
&&&&0.583500&1&broken $z$ translation $(R_{btr,z})$&$A_{1u}$\\
0.52&2&radiative mode&$E_{2g}$&&&&\\
&&&&&\\
0.75&1&breathing mode&$A_{1g}$&0.76&1&breathing mode&$(A_{1g})$\\
0.84&2&a dipole `breathing' motion\tnote{$\dagger$}&$E_{1u}$&0.846727&2&
               broken $x$, $y$ translation $(R_{btr,x}, R_{btr,y})$&$E_{1u}$\\
&&&&&\\
\bottomrule
\end{tabular}
\caption{Comparison of vibrational frequency between the numerical
results and the rational map ansatz, $B=2$.}\label{table4}
\begin{tablenotes}
\item [$\dagger$] Its description of the energy density matches that of the 
broken $x$, $y$ translational zero mode obtained from the rational map ansatz; 
however, this representation, $E_{1u}$, coincides with the doublet of $x$, $y$ 
translational zero mode of the 2-monopole toroidal BPS solution.
\end{tablenotes}
\end{threeparttable}
\end{center}
\end{sidewaystable}

\begin{sidewaystable}
\begin{center}
\begin{threeparttable}\fontsize{7}{12pt}\selectfont
\begin{tabular}{cclc|cclc|cc}
\toprule
&&&&&&\\
&&Numerical\cite{BBT1}&&&&Ansatz\\
&&&&&&\\
\midrule
&&&&&&\\
$\omega$&Degeneracy&Description&Symmetry&$\omega$&Degeneracy&Description\\
&&&&&&\\
&&&&0&2&iso-rotation $(R_{iso,x}, R_{iso,y})$&$E_u$&$E^O$&2\\
0.070&3&broken rotation&$F_{1g}$&0&3&rotation $(R_{rot,x}, R_{rot,y}, R_{rot,z})$\\
0.367&2&deuteron scattering mode&$E_g$&1.006370&2&
            deuteron scattering mode $(R_{scat1}, R_{scat2})$\\
0.405&1&tetrahedral mode&$A_{2u}$&1.402608&1&tetrahedral mode $(R_{tet})$\\
0.419&3&rhombus mode&$F_{2g}$&0.911260&3&rhombus mode 
          $(R_{rhomb,x}, R_{rhomb,y}, R_{rhomb,z})$\\
0.513&3&4 Skyrmions mode&$F_{2u}$&0.750841&3&4 Skyrmions mode 
        $(R_{tb,x}, R_{tb,y}, R_{tb,z})$\\
&&&&&&\\
0.545&2&radiative mode&$E_g$&&&\\
&&&&0&1&iso-rotation $(R_{iso,z})$\\
0.587&1&radiative mode&$A_{2u}$&&&\\
0.605&1&breathing mode&$A_{1g}$&0.64&1&breathing mode\\
0.655&3&broken translation&$F_{1u}$&0.562538&3&broken translation 
          $(R_{btr,x}, R_{btr,y}, R_{btr,z})$\\
0.738&3&diagonal mode&$F_{2g}$&&\\
0.908&3&lowest nonzero radiative mode&&&\\
&&&&&&\\
\bottomrule
\end{tabular}
\caption{Comparison of vibrational frequency between the numerical
result and the rational map ansatz, $B=4$.}\label{table5}
\begin{tablenotes}
\item [$\dagger$] Its description of the energy density matches that of the 
broken $x$, $y$ translational zero mode obtained from the rational map ansatz; 
however, this representation, $E_{1u}$, coincides with the doublet of $x$, $y$ 
translational zero mode of the 2-monopole toroidal BPS solution.
\end{tablenotes}
\end{threeparttable}
\end{center}
\end{sidewaystable}

\section{Conclusion}
The rational map ansatz has been shown to be a good approximation to
the shell-like solutions of the Skyrme model, {\it i.e.} when the pion mass 
is null or relatively small, or when the baryon number is small. In this paper 
we have shown that the rational map ansatz can also be used to study the 
vibration modes of the Skyrmion solutions and that qualitatively it produces 
the correct vibration modes. For $B=1$, for which the ansatz is actually the 
exact solution,  and $B=2$, the vibration modes are all predicted, but for 
$B=4$ they are all predicted except for one, the one with the largest 
vibration frequencies.

On the other hand, the vibration frequencies obtained from the rational map 
ansatz are quite poor when compared to the frequencies obtained numerically
in \cite{BBT1} and \cite{BBT2}. The relative order is not even correct.
We interpret this result by saying that the rational map ansatz is too stiff
and that the configuration is restrained to vibrate in a subspace that is 
too narrow.

Our study has also shown that some of the modes observed by Barnes et al. were 
mistakenly taken as genuine vibration modes when they are actually broken 
vibration modes. 

%\section{Acknowledgement}
\vskip 5mm
{\Large \bf Acknowledgement}

We would like to thanks P. Sutcliffe for useful discussions and comments.

%%%%%%%%%%%%%%%%%%%%%%%%%%%%%%%%%%%%%%%%%%%%%%%%%%%%%%%%%%%%%%%%%%%%%%%%%%%%

\end{document}